\documentclass[]{spie}  %>>> use for US letter paper
\usepackage{amsmath,amsfonts,amssymb}
\usepackage[]{graphicx}
\usepackage[colorlinks=true, allcolors=blue]{hyperref}

\title{XPipeline: Starlight subtraction at scale for MagAO-X}

\author[a]{Joseph~D.~Long}
\author[a]{Jared~R.~Males}
\author[a]{Sebastiaan~Y.~Haffert}
\author[a]{Laird~M.~Close}
\author[a]{Katie~M.~Morzinski}
\author[a]{Kyle~Van~Gorkom}
\author[b]{Jennifer~Lumbres}
\author[b]{Warren~Foster}
\author[b]{Alexander~Hedglen}
\author[b]{Maggie~Kautz}
\author[b]{Alex~Rodack}
\author[c]{Lauren~Schatz}
\author[c]{Kelsey~Miller}
\author[d]{David~Doelman}
\author[d]{Steven~Bos}
\author[d]{Matthew~A.~Kenworthy}
\author[d]{Frans~Snik}
\author[e]{Gilles~P.~P.~L.~Otten}

\affil[a]{Steward Observatory, University of Arizona, Tucson, 933 N. Cherry Ave., Tucson, AZ 85721, USA}
\affil[b]{Wyant College of Optical Sciences, University of Arizona, Tucson, 1630 E. University Blvd., Tucson, AZ 85721, USA}
\affil[c]{Air Force Research Lab, Kirtland Air Force Base, New Mexico, USA}
\affil[d]{Leiden Observatory, Leiden University, P.O. Box 9513, 2300 RA Leiden, The Netherlands}
\affil[e]{Academia Sinica, Institute of Astronomy and Astrophysics, 11F Astronomy-Mathematics Building, NTU/AS campus, No. 1, Section 4, Roosevelt Rd., Taipei 10617, Taiwan}
\authorinfo{Send correspondence to J.D.L.: josephlong@arizona.edu.}
\begin{document}
  \maketitle

%%%%%%%%%%%%%%%%%%%%%%%%%%%%%%%%%%%%%%%%%%%%%%%%%%%%%%%%%%%%%
\begin{abstract}
  MagAO-X is an extreme adaptive optics (ExAO) instrument for the Magellan Clay 6.5-meter telescope at Las Campanas Observatory in Chile. Its high spatial and temporal resolution can produce data rates of 1 TB/hr or more, including all AO system telemetry and science images. We describe the tools and architecture we use for commanding, telemetry, and science data transmission and storage. The high data volumes require a distributed approach to data processing, and we have developed a pipeline that can scale from a single laptop to dozens of HPC nodes. The same codebase can then be used for both quick-look functionality at the telescope and for post-processing. We present the software and infrastructure we have developed for ExAO data post-processing, and illustrate their use with recently acquired direct-imaging data.
\end{abstract}

%>>>> Include a list of keywords after the abstract

\keywords{coronagraphy, post-processing, high-performance computing}

%%%%%%%%%%%%%%%%%%%%%%%%%%%%%%%%%%%%%%%%%%%%%%%%%%%%%%%%%%%%%
\section{INTRODUCTION}
\label{sec:intro}

Extreme adaptive optics, or ExAO, pushes the limits of adaptive optics to enable detection and characterization of faint companions at small separations from their host stars. The MagAO-X instrument at the Magellan Clay 6.5-meter telescope in Chile is designed to correct for atmospheric aberrations at 2 kHz with a pyramid wavefront sensor (PWFS) and three deformable mirrors (DMs)\cite{malesMagAOXFirstLight2020}. The wavefront sensor and deformable mirror data streams provide information on the state of the system beyond what is captured in the science focal planes. This may prove useful for starlight subtraction, but scaling conventional starlight subtraction algorithms to very large numbers of observations (or high-dimensional observations, or both) quickly becomes computationally intractable due to the scaling behavior of large matrix decompositions \cite{LongMales2021}. Nevertheless, MagAO-X is designed to save every bit of information collected with the observation, and observers can access it for use in their analysis.

In order to enable efficient analysis of data from extreme adaptive optics instruments, we have prototyped our instrument pipeline on top of distributed execution platforms which allow mostly-unchanged Python code to distribute computations over multiple CPU cores and even multiple machines in a cluster configuration. This work describes the movement of data and commands through the MagAO-X system and the XPipeline post-processing architecture.

\section{BACKGROUND}
\label{sec:background}

\begin{figure}
  \begin{center}
  \begin{tabular}{c}
  \includegraphics[width=0.99\textwidth]{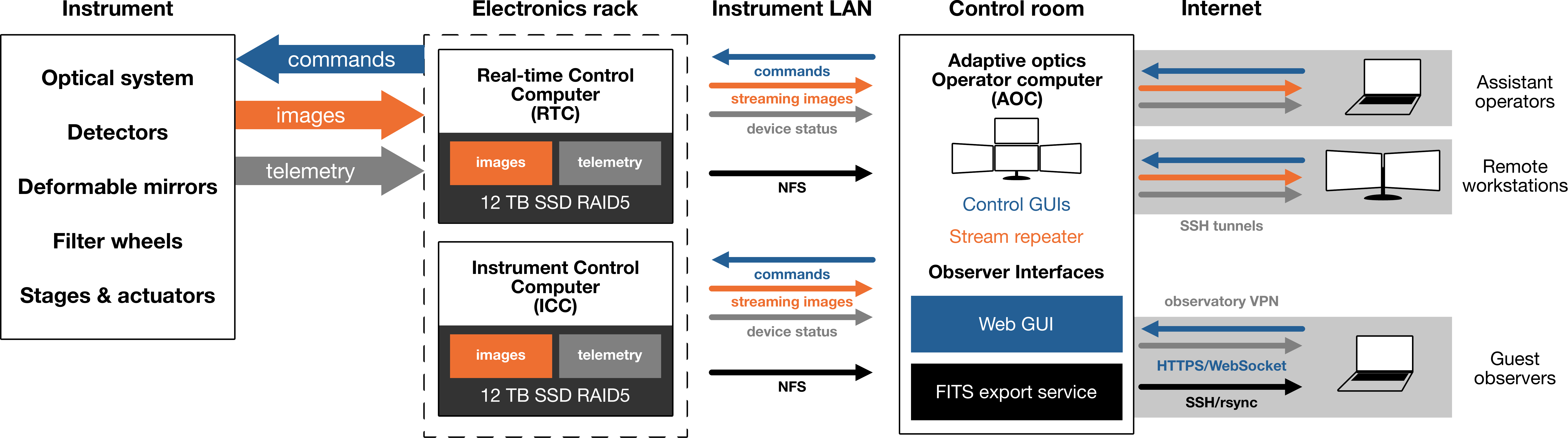}
  \end{tabular}
  \end{center}
  \caption[dataflow]
%>>>> use \label inside caption to get Fig. number with \ref{}
  { \label{fig:dataflow} Command and data flow within MagAO-X, organized with the instrument hardware on the left and control computers on the right. The adaptive optics control computer (AOC) runs control GUIs locally, but also serves as the home for guest observer services like reduced-frame-rate image streams, web GUIs, and the service that produces science-ready FITS files for observers.}
\end{figure}

The MagAO-X instrument is described in terms of its light path and its control loops elsewhere in these proceedings, but here we provide an overview of the path of machine-readable commands into the system and data streams out of the system. The diagram in Figure~\ref{fig:dataflow} illustrates these pathways, and how they enable remote operation and a loose coupling of user-interface code to the actual hardware control functionality.

%-------------
%-------------

\subsection{Commanding and device status}

The commanding of the MagAO-X instrument's many controllable degrees of freedom, with the notable exception of latency-sensitive deformable mirror positions, is accomplished with the Instrument-Neutral Distributed Interface, or INDI\cite{indiProtocol}. The INDI protocol enforces an architecture in which all controllable instrument properties are described by a {\tt deviceName.propertyName.elementName} triple. Each property is either a number, a text string, a switch, or a light (status indicator) with metadata like the valid limits (for a number) or the number of simultaneously active elements allowed (for a switch). Client applications receive updates to properties they indicate an interest in, and can send new values (which the receiving device process may or may not choose to accept).

The protocol notably does not enforce access controls, leaving that to the application. In our system, all INDI communication uses sockets bound to {\tt localhost}, available only to software running on the same computer or over SSH tunnels (which provide authentication and authorization).

The properties of interest to the AO operator are largely exposed through C++/Qt GUIs that run on the operator workstation. However, since simple TCP sockets carry all commands, these can be used remotely on any computer with the software installed. The software system for MagAO-X can be installed in an automated fashion in a virtual machine (VM) which then connects to the instrument like any other workstation\cite{longRemoteOperation}. This remote operation capability proved invaluable during the pandemic when occupancy limits meant not everyone who needed to use the instrument could access the lab or telescope control room.

\begin{figure}
  \begin{center}
  \begin{tabular}{c}
  \includegraphics[width=\textwidth]{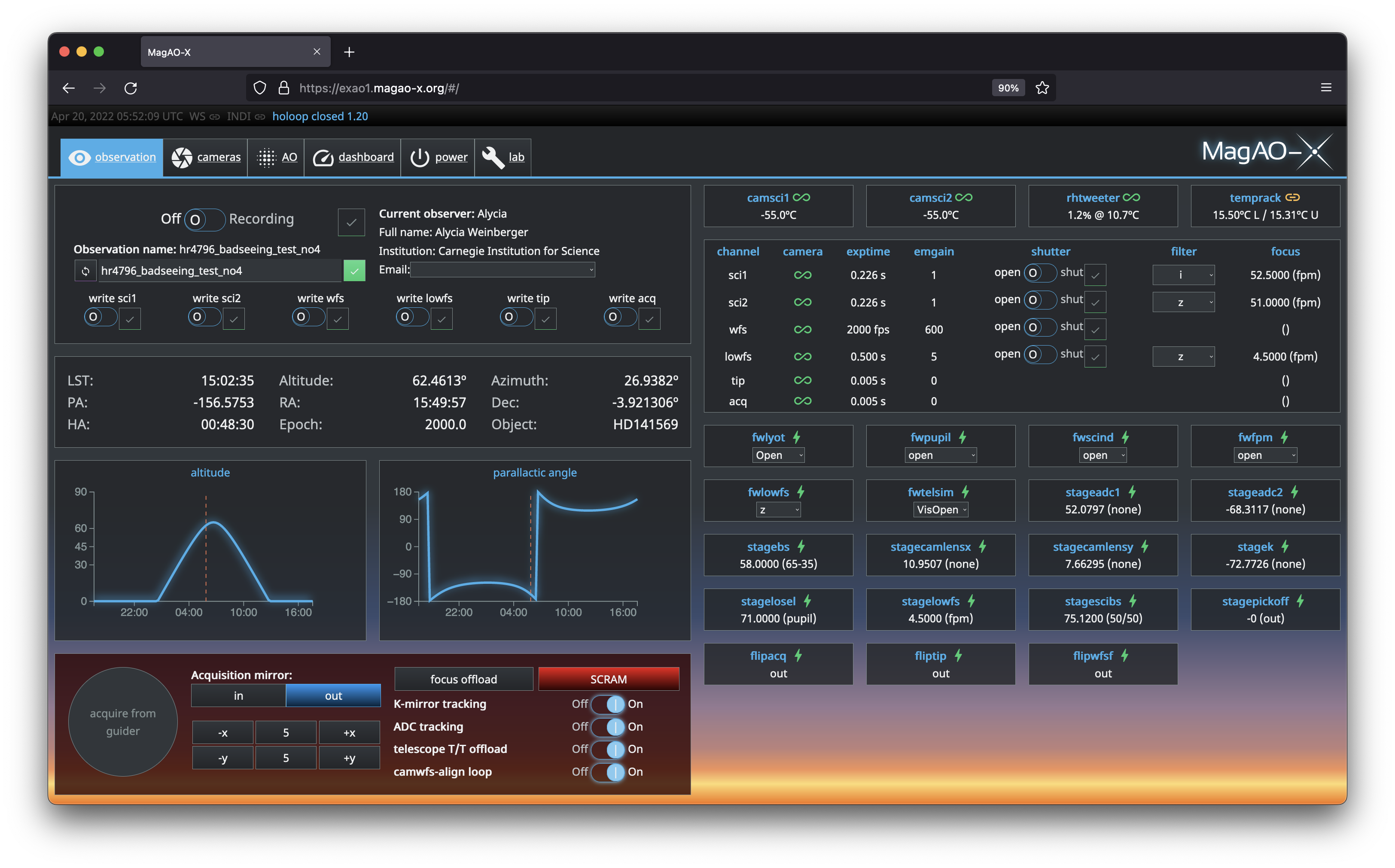}
  \end{tabular}
  \end{center}
  \caption[dataflow]
%>>>> use \label inside caption to get Fig. number with \ref{}
  { \label{fig:webui} The MagAO-X web interface is designed to balance flexibility with logical organization, so that new capabilities added to the instrument require minimal bespoke code to integrate in the web interface. When attached to the Magellan telescope, telescope control system (TCS) information is replicated to the web interface for guest observers and AO operators who cannot see the telescope operator's screen. The web UI also includes some sanity checks to warn operators if their instrument configuration would do something they likely did not intend.}
\end{figure}

To expose a subset of properties for guest observers and operators, the MagAO-X web GUI application wraps them in a web-based interface that is simpler for guest observers to access without installing additional software or creating a VM. This interface is shown in Figure~\ref{fig:webui}. Since the user's browser cannot communicate directly with the INDI server (and devices), a Python server process translates commands to and from the INDI protocol format. The resulting instrument state is connected to UI controls with Vue, a JavaScript library for building application interfaces.

No special treatment is required to ensure properties set in Qt GUIs are updated in the web interface, or vice versa. Likewise, the system does not need to distinguish between commands from a physically-present operator sitting at the control computer and those from a remote operator.

\subsection{Streaming}

Low-latency communication is accomplished through shared-memory images (``shmims'') provided as part of the CACAO and MILK packages\cite{cacao2018,cacao2020} which implement synchronization primitives around a ring buffer. This data structure allows the AO system to synchronize access to things like wavefront sensor images and DM commands. MagAO-X also uses these images to hold science focal plane images as they are read out. Persisting science data and AO data to disk is then accomplished identically, with a MagAO-X software process that reads these shared memory images and writes them to a compressed format on the computer's local RAID5 SSD array.

As the real-time control computer (RTC) may need to access images from the instrument control computer (ICC), and the adaptive optics operator computer (AOC) needs to visualize both, it is necessary to replicate these data over the network. A helper application watches for updates and transmits them over the instrument's local area network (LAN) to clients which write them into shared-memory images on the receiving computer (``streaming images'' in Figure~\ref{fig:dataflow}). Remote operation uses a special mode of this same helper application that limits the maximum frame rate, as even compressed data can saturate a consumer internet connection (``stream repeater'' in Figure~\ref{fig:dataflow}).

\subsection{Storage and access}

For maximum throughput, image data (including DM commands and WFS images) are stored to a RAID5 array of SSDs attached to the computer (ICC or RTC) where they are acquired or computed. Compression with the eXtreme Reordered Image Format (XRIF) library\cite{malesXrif2021} is applied to the stream, achieving 2:1 or better size reduction on integer image data.

Instrument telemetry like filter wheel states and camera temperatures is stored in a binary log format adjacent to the image data. To obtain final images with associated metadata, the XRIF images have to be extracted with access to the telemetry files from both RTC and ICC. To accomplish this, the RTC and ICC data arrays are mounted with NFS on the operator computer. A system service ({\tt lookyloo}) watches for active observations (when an observer name and title are set, and the recording property is ``on'') and extracts the image archives as they become available using the MagAO-X {\tt xrif2fits} utility. These science-ready images are exposed to guest observers with SSH access via rsync.

\section{Distributed pipeline architecture}

MagAO-X can produce thousands of images per second, with observations encompassing many GB or even TB of data. To enable observers to continue to carry out quick-look data reductions on their own computers, while also enabling scalability to larger computing facilities, we implemented a pipeline atop the Ray distributed execution framework\cite{moritzRayDistributedFramework}.

\begin{figure}
  \begin{center}
  \begin{tabular}{c}
    
  \includegraphics[width=0.9\textwidth]{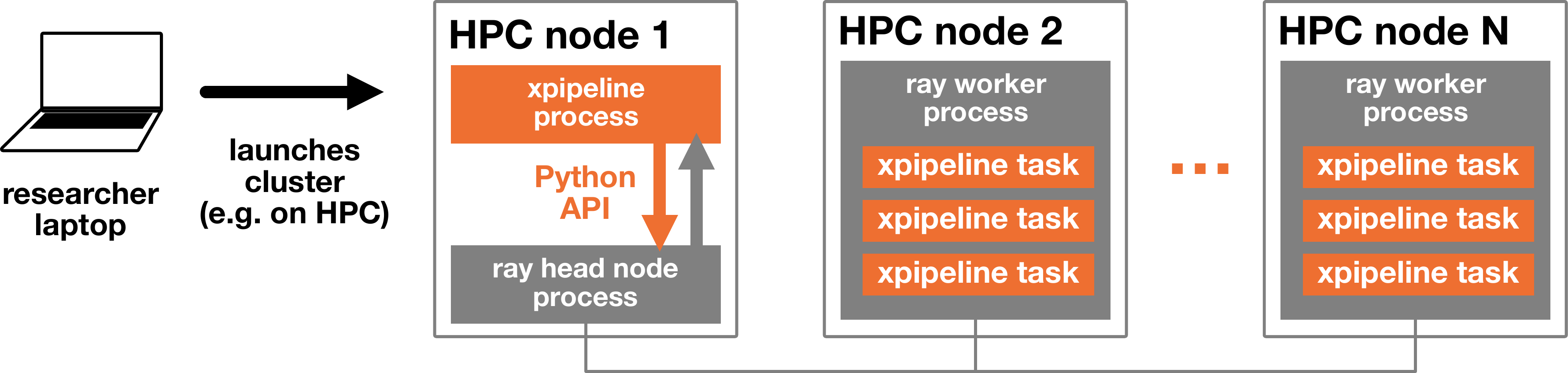}
  \end{tabular}
  \end{center}
  \caption[ray-hpc]
  %>>>> use \label inside caption to get Fig. number with \ref{}
  { \label{fig:ray-hpc} Illustration of the operation of XPipeline on an HPC cluster. The parent xpipeline process on HPC node 1 invokes separable subtasks within the pipeline as Ray ``tasks'', which delegates responsibility for scheduling them and retrieving their outputs to the Ray head node process. Ray workers contact the head node for assignments, and retrieve task arguments over the network. The results are stored in a distributed fashion, to be retrieved by the controlling xpipeline process or a downstream task. The xpipeline process uses the object references obtained from Ray at submission to await the final results of the tasks, which are converted back into plain Python objects for manipulation in the parent process.\\
  \\}
\end{figure}

In single-machine mode, parallel parts of the reduction pipeline (e.g. hyperparameter searches) can be run in parallel on different CPU cores. In cluster mode, the same code can be run across multiple computers, with Ray handling the serialization and deserialization of data over the network between them (Figure~\ref{fig:ray-hpc}).

To validate our architecture, we used a 10,000 frame data cube taken with the gvAPP-180 coronagraph on Clio with MagAO\cite{Otten2017}. This is a very finely temporally sampled dataset for Clio, but still a fraction of the number of frames generated in a typical MagAO-X observation. The data acquired in April 2022 and the upcoming December 2022 MagAO-X observing run will further test the limits of our architecture.

\subsection{Distributed execution on HPC cluster resources}

The task we chose to validate the architecture was a hyperparameter optimization grid for a modified version of the Karhunen-Lo\`eve Image Projections (KLIP) starlight subtraction algorithm. Using 19 nodes with 94 accessible cores each (1,786 total CPU cores), we applied XPipeline in cluster mode to the hyperparameter optimization task. This involved applying the data reduction algorithm to the full data cube injected with a fake signal at 167 locations in the focal plane and four contrast levels (plus the no-injection case), then measuring the recovered SNR at 11 different numbers of modes subtracted to determine the optimal hyperparameters for a blind search. This test was executed on The University of Arizona's Puma high-performance computing (HPC) cluster.

\begin{figure}
  \begin{center}
  \begin{tabular}{c}
  \includegraphics[width=0.5\textwidth]{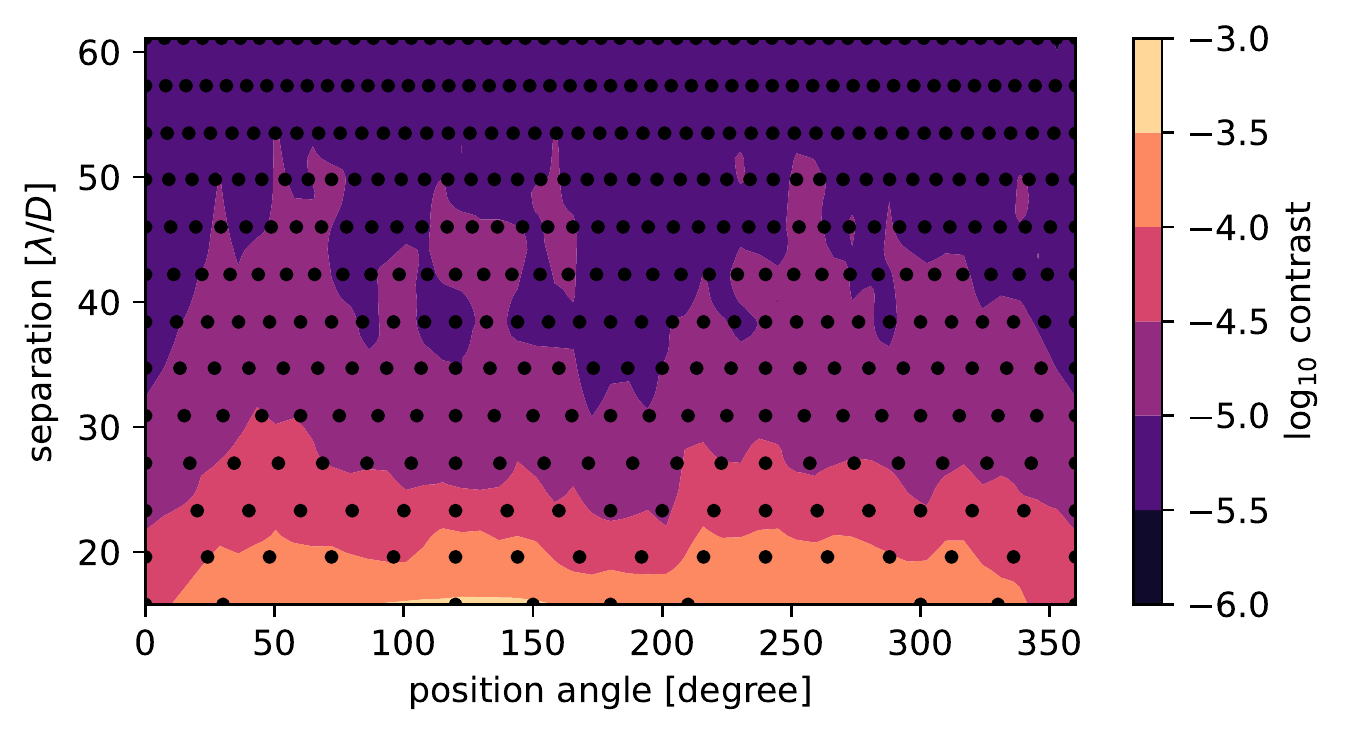}
  \end{tabular}
  \end{center}
  \caption[pa-variation]
%>>>> use \label inside caption to get Fig. number with \ref{}
  { \label{fig:pa-variation} The calibrated contrast floor for an SNR=5 signal as a function of position angle and separation, shown in unwrapped rectilinear coordinates after interpolation and contour-calculation. The focal plane sampling (black points, also shown in Figure~\ref{fig:best-trials}) gets finer in PA as separation increases. The final contrast surface exhibits clear structure with two maxima in contrast as a function of PA.}
\end{figure}

\begin{figure}
  \begin{center}
  \begin{tabular}{c}
  \includegraphics[width=0.9\textwidth]{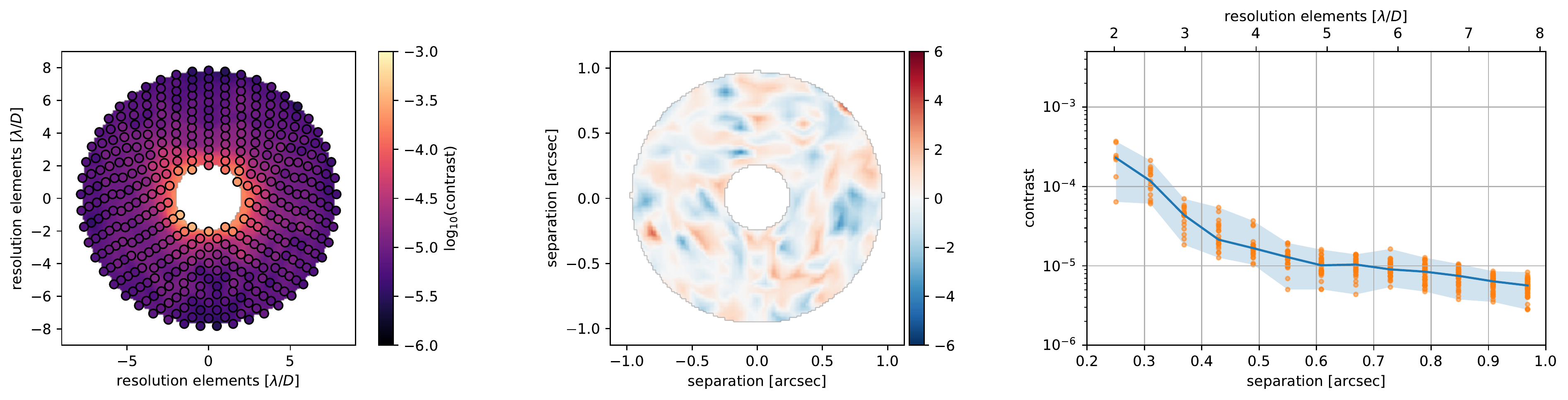}
  \end{tabular}
  \end{center}
  \caption[best-trials]
%>>>> use \label inside caption to get Fig. number with \ref{}
  { \label{fig:best-trials} (left) The calibrated contrast floor for an SNR=5 signal in the example gvAPP-180 data as a function of position in the final derotated focal plane. (middle) The SNR measured at each point with the optimal hyperparameters for that focal plane location (without any injected signal). (right) The calibrated contrast curve as a function of separation, with the shaded region encompassing the variation with position angle.}
\end{figure}

The gvAPP-180 coronagraph forms two complementary dark-hole regions, which means derotated images will have a varying amount of coverage from one or both dark-hole regions. The contrast limits of direct imaging data are expected to vary strongly with separation, but Figure~\ref{fig:pa-variation} illustrates that it varies with position angle as well in these data. As a result, the optimal hyperparameters for recovering a faint companion may not be the same for a whole annulus (as in annular PCA starlight subtraction), but rather be specific to individual focal plane locations. In Figure~\ref{fig:best-trials}, the focal plane locations sampled are shown in the derotated frame, along with the SNR of the detection map and the predicted contrast limits for each point grouped by separation.

\subsection{High-throughput computing with the Open Science Grid}

\begin{figure}
  \begin{center}
  \begin{tabular}{c}
  \includegraphics[width=0.9\textwidth]{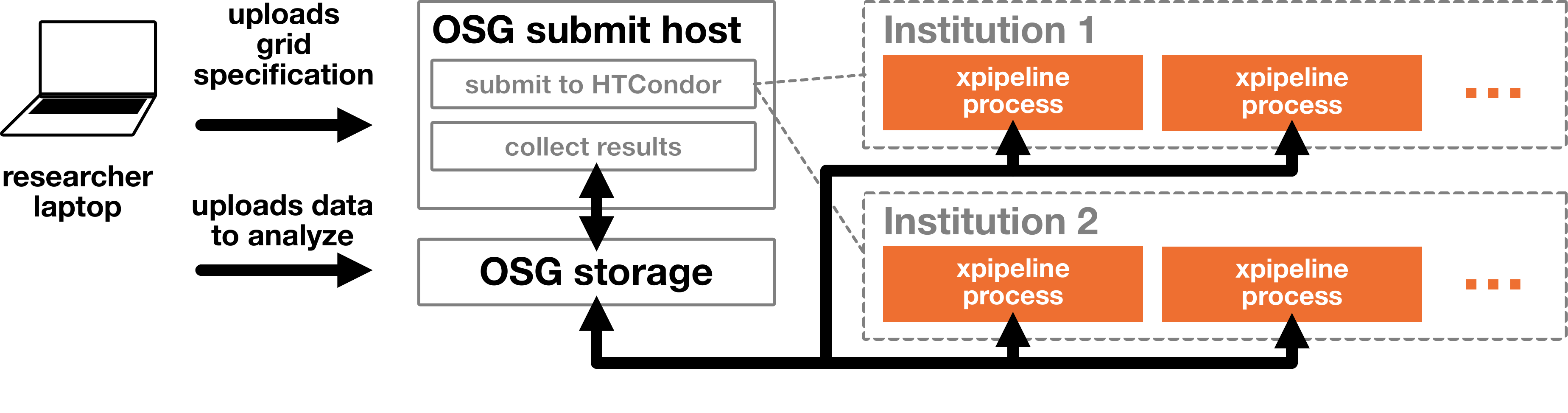}
  \end{tabular}
  \end{center}
  \caption[ray-osg]
%>>>> use \label inside caption to get Fig. number with \ref{}
  { \label{fig:ray-osg} Illustration of the operation of XPipeline on the Open Science Grid. Rather than farming the tasks out through a central coordinator process, the task arguments are determined before submission and supplied to the xpipeline process through HTCondor. This uses the grid framework in xpipeline, but does not necessarily leverage Ray.}
\end{figure}

The Open Science Grid (OSG)\cite{osg07,osg09} is a free resource for US-based researchers that employs HTCondor for high-throughput distributed computing. For tasks like a hyperparameter search, where each task is independent and inputs are known ahead of time, the xpipeline grid tasks can be launched as multiple fully independent worker processes and store their results separately. This use case for the xpipeline package is illustrated in Figure~\ref{fig:ray-osg}. Using a Singularity container, the same execution environment can be made available on many different research computing clusters that donate excess capacity to the Open Science Grid. We successfully employed xpipeline on OSG for an experiment in Bayesian optimization of hyperparameters, about which we hope to share more in future work.

\section{CONCLUSIONS}
\label{sec:conclusions}

The MagAO-X instrument control and data systems provide a wealth of information on instrument performance, and, by persisting it all for later analysis, open the door to telemetry-informed starlight subtraction. However, data volumes are increasing faster than the capacity of individual researchers' laptops. To develop the next generation of ExAO pipelines, we will need to leverage new technologies to distribute computations and handle data sets larger than any single computer's available RAM.

To post-process ExAO data, we must attack the scalability problem from both the algorithmic and the architectural angles. The Ray distributed execution framework provides a radically simplified API for distributing computations from Python code, preserving researcher flexibility for quick iteration and extension to new algorithms. We have not yet reached a point where we are bottlenecked on throughput by the underlying platform, and anticipate RAM usage as our biggest limiting factor going forwards. The Open Science Grid provides a wealth of free high-throughput computing resources, but scaling to the next order of magnitude in data volume may push us beyond their preferred task sizes for efficient scheduling.

%%%%%%%%%%%%%%%%%%%%%%%%%%%%%%%%%%%%%%%%%%%%%%%%%%%%%%%%%%%%%
\acknowledgments     %>>>> equivalent to \section*{ACKNOWLEDGMENTS}

This work has been supported by the Heising-Simons Foundation award \#2020-1824 and NSF MRI Award \#1625441 (MagAO-X). Parts of this research were done using services provided by the OSG Consortium\cite{osg07,osg09}, which is supported by the National Science Foundation awards \#2030508 and \#1836650. S.Y.H. is supported by NASA through the NASA Hubble Fellowship grant \#HST-HF2-51436.001-A awarded by the Space Telescope Science Institute, which is operated by the Association of Universities for Research in Astronomy, Incorporated, under NASA contract NAS5-26555. This work also made use of High Performance Computing (HPC) resources supported by the University of Arizona.

%%%%%%%%%%%%%%%%%%%%%%%%%%%%%%%%%%%%%%%%%%%%%%%%%%%%%%%%%%%%%
%%%%% References %%%%%

\bibliography{report}   %>>>> bibliography data in report.bib
\bibliographystyle{spiebib}   %>>>> makes bibtex use spiebib.bst

\end{document}